\documentclass[10pt,letterpaper,twocolumn]{article} 

\usepackage{ol}
\usepackage{hyperref}
\usepackage{amsmath}

\begin{document}

\twocolumn[ 

\title{Transportable laser system for atom interferometry}

\author{P. Cheinet, F. Pereira Dos Santos, T. Petelski, J. Le Gou\"et, K. T. Therkildsen,\\
A. Clairon and A. Landragin}

\address{LNE-SYRTE, CNRS UMR8630, Observatoire de Paris, 61 avenue de l'observatoire, 75014 Paris,
France}




\begin{abstract}
We describe an optical bench in which we lock the relative frequencies or phases of a set of three
lasers in order to use them in a cold atoms interferometry experiment. As a new feature, the same
two lasers serve alternately to cool atoms and to realize the atomic interferometer. This requires
a fast change of the optical frequencies over a few GHz. The number of required independent laser
sources is then only 3, which enables the construction of the whole laser system on a single
transportable optical bench. Recent results obtained with this optical setup are also presented.
\end{abstract}


] 

\newpage


\noindent Within the last decades, atom interferometers have developed into a highly competitive
tool for precision measurements.\cite{Bordé} Atomic fountains used as atomic clocks are the best
realization of the time unit.\cite{Clock} Atom interferometry also promises sensors to be highly
sensitive to inertial forces.\cite{Riehle,Pritchard97,Kasevich00,Peters01} The use of stimulated
Raman transitions to manipulate the atomic wave packet has proven to be an efficient way to obtain
high accuracy devices.\cite{Kasevich00,Peters01}

In this letter, we describe a robust, compact and versatile laser system for atom interferometers
using alkali atoms. Such experiments basically need two different optical frequencies, whose
difference remains close to the hyperfine transition frequency. When they are tuned close to the D2
transitions, they are used to cool and repump the atoms in a magneto-optical trap (MOT). When far
detuned, and phase locked, they are used to induce stimulated Raman transitions for the
interferometer.\cite{Kasevich91} Since lasers are not used simultaneously for trapping and Raman
transitions, we have implemented a technique to use the same two lasers for both functions on our
gravimeter.\cite{IEEE} It allowed us to build the whole laser setup on a $60\times90$ cm$^2$
optical bench.

\begin{figure}[htb]
\centerline{\includegraphics[width=8cm]{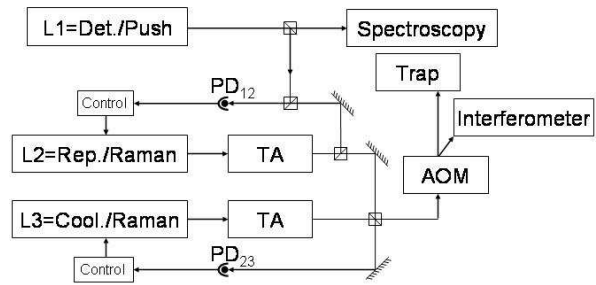}} \caption{Laser setup. The detection laser is
locked on a spectroscopy signal. The repumper (resp. cooling) laser frequency is compared to the
detection (resp. repumper) laser frequency with an optical beat note and servo locked. Two tapered
amplifiers (TA) are used on the repumper and cooling lasers before they are combined on a
polarizing beam splitter cube and sent alternately to the trap or to the interferometer using an
acousto-optical modulator (AOM).}\label{principe}
\end{figure}

Our laser setup is shown in figure \ref{principe}. A first laser L1 is locked on an atomic
transition, using FM-spectroscopy\cite{Hall81} on a saturated absorption signal. This laser
constitutes an optical frequency reference and is used in our experiment to detect or push the
atoms. A second laser L2 is alternately used as repumper or as master Raman laser. Part of the
outputs of L1 and L2 are superimposed on a fast photodetector (PD$_{12}$) (Hamamatsu G4176) and the
frequency of the beat note is servo locked by using a frequency to voltage converter. A third laser
L3 is used alternately as cooling or as slave Raman laser. The frequency difference between L2 and
L3 is measured with a second optical beat note on PD$_{23}$. Finally, both L2 and L3 beams are
superimposed and directed through an acousto-optical modulator either to a magneto-optical trap or
to an atomic interferometer.

\begin{figure}[htb]
\centerline{\includegraphics[width=8cm]{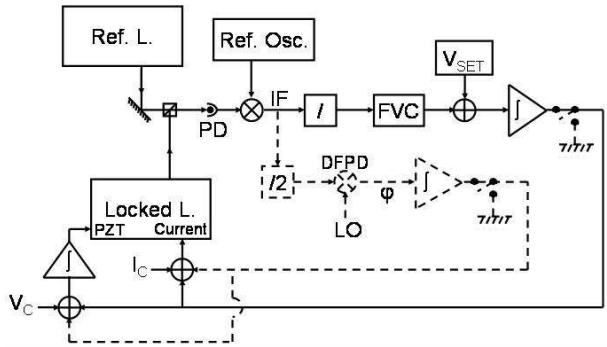}} \caption{Locking electronics. (Solid
line) Frequency lock scheme. The optical beat note is mixed with a reference oscillator to an
intermediate frequency (IF). The IF is converted to a voltage signal and another voltage $V_{Set}$
is subtracted to obtain the error signal of the lock. It is integrated and then sent to the current
driver. It is integrated once more and sent to the PZT driver. Predetermined corrections $I_C$ and
$V_C$ are added to the diode current and to the PZT voltage during the sweep. (Dotted line) Phase
lock scheme added to L3. The IF is compared to a Local Oscillator (LO) in a digital phase and
frequency detector (DPFD) delivering the phase error signal. Two switches select which loop is
closed.} \label{elec}
\end{figure}

Both frequency locks of L2 and L3 use the same scheme which is shown in figure \ref{elec}. The
optical beat note issued from the photodetector is mixed with a reference oscillator, down to an
intermediate frequency (IF). For the L2 lock this reference is a tunable oscillator (YIG). Whereas
for L3 we use a fixed 7 GHz frequency obtained by the multiplication of a low phase noise 100 MHz
quartz oscillator. The IF signal is then sent into a digital frequency divider in order to fit into
the working frequency range (0-1 MHz) of a frequency to voltage converter (FVC) (AD650). A computer
controlled offset voltage $V_{Set}$ is subtracted from the output voltage of the FVC. The obtained
error signal is integrated once and added to the laser diode current. This correction signal is
integrated again and added to the piezoelectric (PZT) voltage which controls the cavity length. To
change the laser frequency, one can change $V_{Set}$ for fine tuning or the YIG frequency for
larger frequency changes. In addition, a computer controlled correction current $I_C$ and a
correction voltage $V_C$, are added to the current and PZT drivers to help the lock while changing
the laser frequency.

For the phase lock of L3 a second path is implemented. The IF frequency is divided by $2$ and
compared, in a digital phase and frequency detector (DPFD)\cite{Tino} (MCH12140), to the signal of
a local oscillator at 82.6 MHz which is generated by a Direct Digital Synthesiser (DDS) (AD9852)
clocked at 300 MHz. The DPFD delivers an error signal which is added through a high bandwidth servo
system ($\sim$ 4 MHz) to the laser current. It is also added to the PZT error signal before its
last integration. Moreover some switches can be activated so that either the frequency lock loop or
the phase lock loop is closed.\\

Our interferometer is an atomic gravimeter which measures the acceleration of freely falling
$^{87}$Rb atoms. Its sensitivity is given by: $\Delta \Phi=k_{eff} g T^2$, where $\Delta \Phi$ is
the interferometric phase, $k_{eff}$ is the effective wave vector of the Raman transition, $g$ is
the Earth's gravity and $T$ is the time between the interferometer's Raman pulses.

This frequency locking system is versatile and enables to control dynamically the frequency of the
two lasers, over the whole experimental sequence. It is first possible to frequency lock the lasers
to the frequencies required to cool $^{87}$Rb atoms in a MOT. Dividing the total available laser
power between a 2D-MOT\cite{Dieckmann98} and a 3D-MOT, loading rates of $3\times10^9$
atoms.s$^{-1}$ are obtained. Then we turn the magnetic field off and further cool the atoms with
$\sigma^+ - \sigma^-$ molasses down to a temperature of 2.5 $\mu$K.

Once the atoms have been released from the molasses, a frequency ramp is applied on the YIG
oscillator. This ramp induces a detuning $\Delta$ of up to 2 GHz on both L2 and L3 to get the Raman
laser frequencies. We also add a ramp on the PZT voltages $V_C$ to induce a 2 GHz sweep so that the
laser frequencies stay inside the locking range. Since the PZT mode-hop free tuning range is close
to $\pm$ 0.6 GHz, it is necessary to change the current setting point of the laser during the
sweep. Thus we apply ramps on the currents $I_C$ so that the laser frequencies remain in the middle
of the free tuning range. When the servo loop is closed, the lasers stay locked during the whole
sequence.

\begin{figure}[htb]
\centerline{\includegraphics[width=8cm]{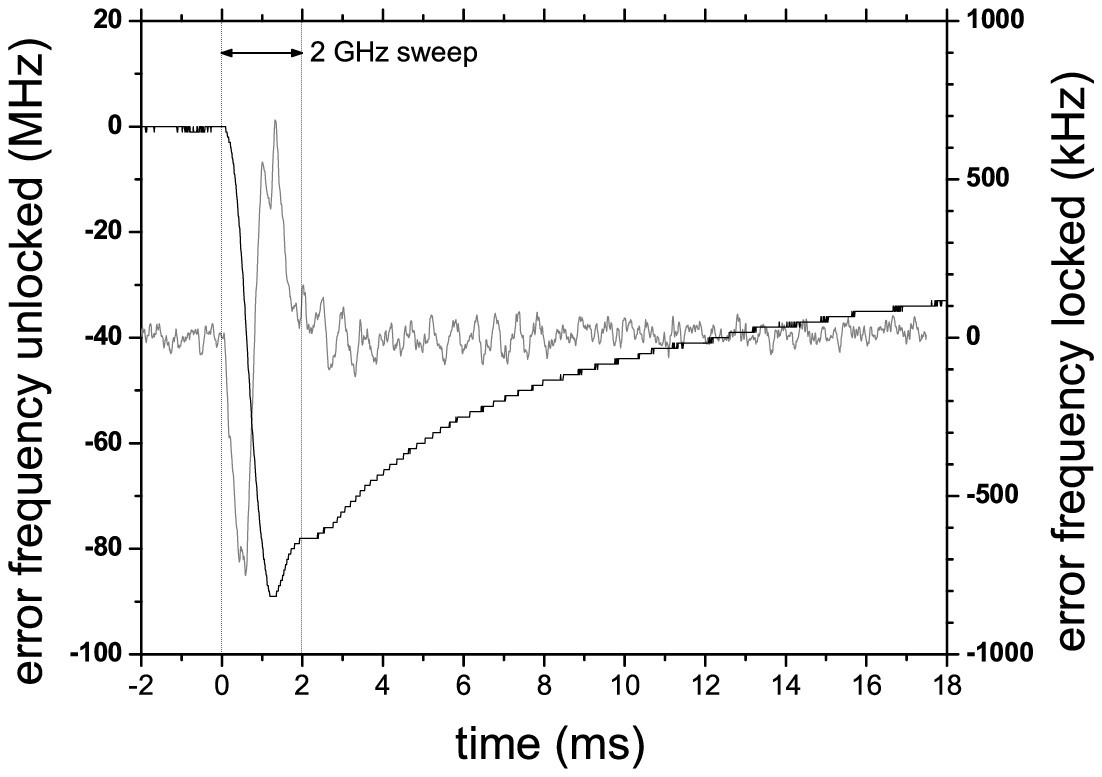}} \caption{L2 frequency error during a 2 GHz sweep
imposed in 2 ms. (Black) Servo loop opened (laser unlocked). (Gray) servo loop closed (laser
locked). }\label{2GHz}
\end{figure}

In figure \ref{2GHz} is shown the response of the servo system to a frequency ramp of 2 GHz in 2
ms, in open and closed loop configurations. The black curve corresponds to the error signal of L2
in open loop operation. The laser frequency remains within 100 MHz from the locking point during
the whole 2 GHz ramp. The voltage ramp does not compensate exactly the sweep because of thermal
effects due to the change in the laser current. When the servo loop is closed, the remaining
frequency deviation is compensated for. The gray curve shows the residual frequency error of L2
during the sweep, and reveals residual damped oscillations of the PZT.

\begin{figure}[htb]
\centerline{\includegraphics[width=8cm]{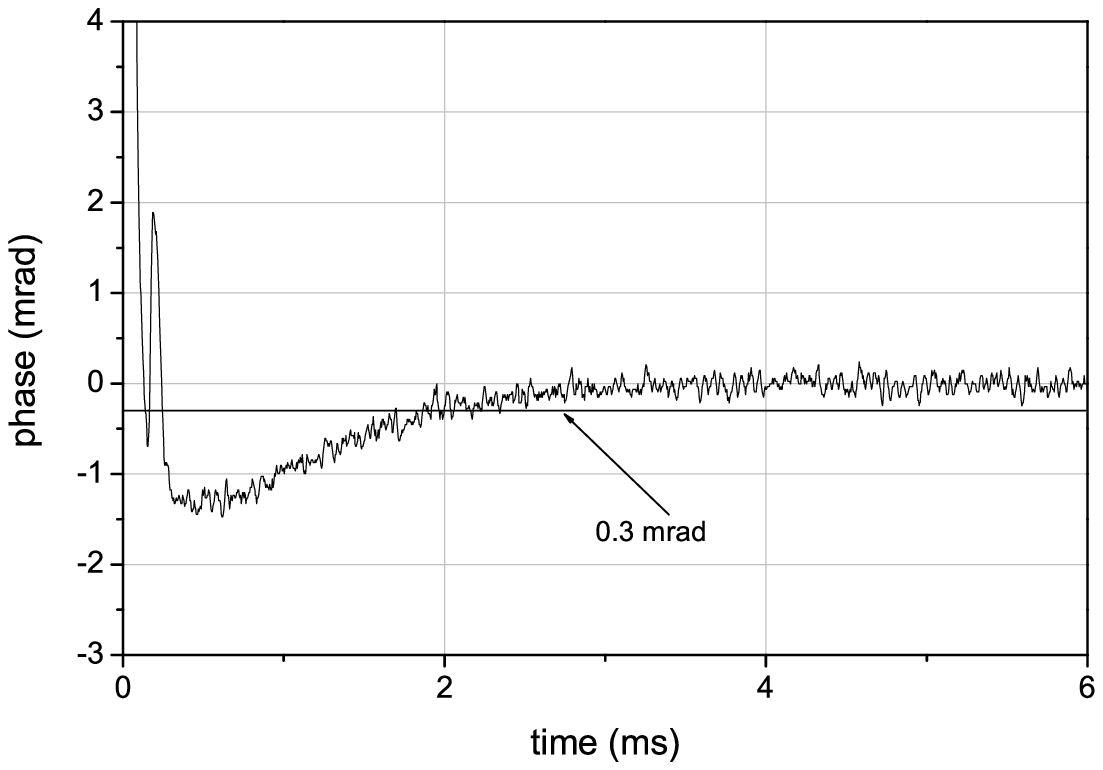}} \caption{L3 phase error. The PLL is closed
at $t=0$ after the 2 GHz sweep. After 0.5 ms, the phase error is exponentially decreasing with a
time constant of 2 ms. }\label{transition}
\end{figure}

We then switch L3 to the phase-locked loop (PLL) after the end of the frequency ramp. We aim at
obtaining an accuracy of $10^{-9}$ $g$ which implies that the phase error has to remain below 0.3
mrad.\cite{IEEE} It takes a few hundreds of $\mu$s for the lock to come perfectly to the right
frequency and to start phase locking. We display the residual phase error as a function of the
delay after enabling the PLL in figure \ref{transition}. 0.5 ms after the loop is closed, the phase
reaches a steady state with a 2 ms time constant exponential decay. The 0.3 mrad criterion is then
reached in about 2 ms. We have measured its spectral phase noise density in steady state\cite{IEEE}
and calculated a total contribution of 0.56 mrad rms of phase noise in the atomic interferometer
corresponding to $10^{-9}$ $g$ rms.

We want to emphasize that the Raman detuning $\Delta$ can be changed at will and other sweeps can
be added in the cycle. This enables to realize first a velocity selective Raman pulse ($\sim$ 35
$\mu$s), with a detuning of 2 GHz which reduces the spontaneous emission. Then the detuning is
swept back to 1 GHz for the interferometer itself, to achieve a better transfer efficiency.

\begin{figure}[htb]
\centerline{\includegraphics[width=8cm]{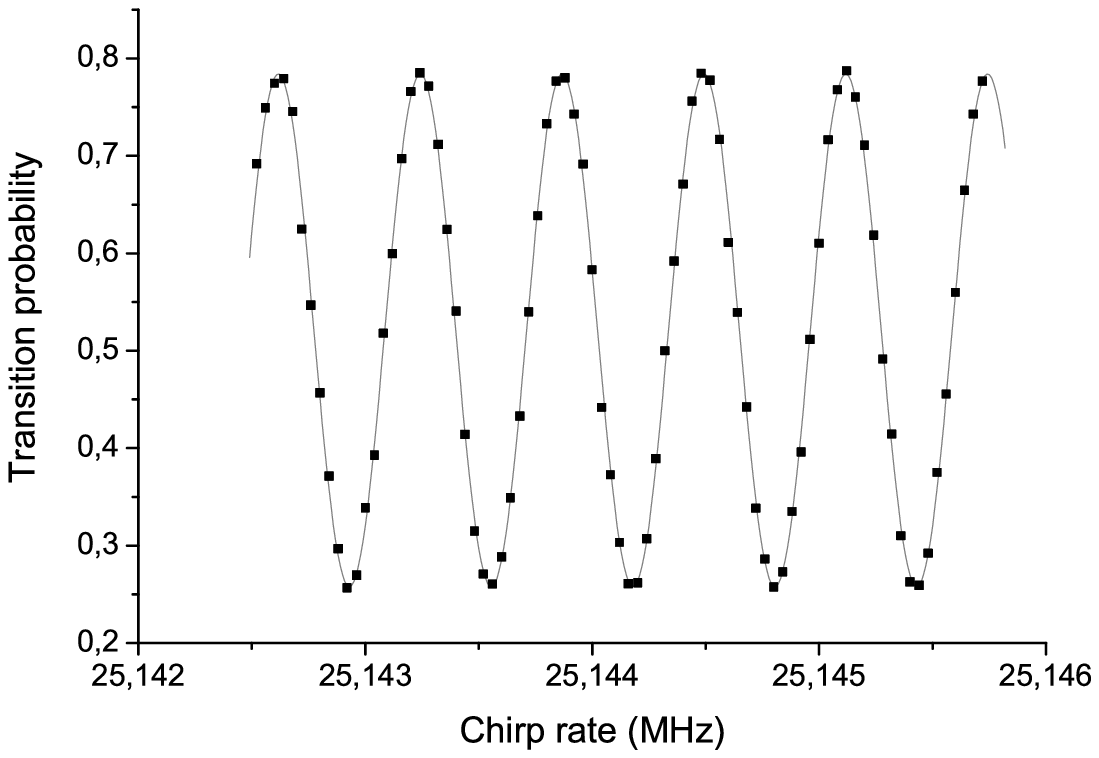}} \caption{Atomic interferometer fringes
obtained by scanning the Raman detuning chirp rate during the interferometer. The time between the
Raman pulses is $T=40$ ms. The solid line is a sinuso\"idal fit of the experimental points
displayed in black squares.}\label{franges}
\end{figure}

Finally, the phased-locked Raman lasers are used to realize the interferometer. Due to the Doppler
effect, the Raman detuning has to be chirped to compensate for the increasing vertical velocity of
the atomic cloud. This chirp $a$, obtained by sweeping the DDS frequency, induces an additional
phase shift. The total interferometric phase is then given by: $\Delta \Phi=(k_{eff} g-a) T^2$.
Figure \ref{franges} displays the interferometric fringes obtained by scanning the chirp rate. In
this experiment, $T$ is 40 ms and the sensitivity is already of $4 \times 10^{-8}$ $g$ Hz$^{-1/2}$,
limited by residual vibrations of the apparatus.

To conclude, this locking technique allowed us to build with only three lasers an optical bench
providing the required frequencies to cool $^{87}RB$ atoms in a 3D-MOT and to realize an atomic
interferometer with far detuned Raman lasers. Our laser setup is robust and versatile since the
lasers routinely stay locked for days and we can change the detuning of the Raman transitions at
will. Our goal for the gravimeter experiment is to reach an accuracy of $10^{-9}$ $g$ and a
sensitivity of a few $10^{-9}$ $g$ Hz$^{-1/2}$. Thanks to its compactness, the gravimeter will be
transportable to compare it with other absolute gravimeters. It will also be moved close to the LNE
watt balance experiment, which aims at measuring the Planck's constant and redefining the
kilogram.\cite{Geneves05}

The authors P. C. and J. L. G. thank DGA for supporting this work. The author K. T. T. thanks also
the "Fondation Danoise" for its support.

\end{document}